\documentclass[pra,aps,twocolumn,10pt,byrevtex,notitlepage,secnumarabic,nofootinbib,longbibliography,a4paper,showkeys,nobalancelastpage,preprintnumbers]{revtex4-1}

\usepackage{graphicx}
\usepackage[utf8]{inputenc}
\usepackage{siunitx}
\usepackage[colorlinks=true,urlcolor=blue,linkcolor=blue,citecolor=blue,breaklinks=true]{hyperref}
\usepackage{breakurl}
\usepackage{natbib}
\usepackage{mathtools}
\usepackage{bibentry}

\AtBeginDocument{\usepackage{booktabs}}

\begin{document}
\author{José María Martín-Olalla}
\email{olalla@us.es}
\date{February 6, 2020}
\thanks{\includegraphics[scale=0.6]{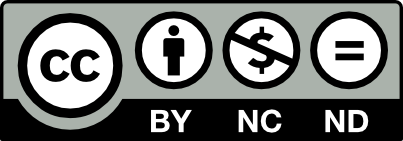}}
\affiliation{Universidad de Sevilla. Facultad de Física. Departamento de Física de la Materia Condensada. ES41012 Sevilla. Spain}
\title{Traffic accident increase attributed to Daylight Saving Time doubled after Energy Policy Act}
\preprint{Current Biology (2020) \textbf{30}(7) R298-R300, doi: \href{https://doi.org/10.1016/j.cub.2020.03.007}{https://doi.org/10.1016/j.cub.2020.03.007} Licences CC-BY-NC-ND}

\begin{abstract}

On January 30, 2020 Current Biology released the report \href{https://doi.org/10.1016/j.cub.2019.12.045}{``A Chronobiological Evaluation of the Acute Effects of Daylight Saving Time on Traffic Accident Risk''} (doi: \href{https://doi.org/10.1016/j.cub.2019.12.045}{10.1016/j.cub.2019.12.045} by Fritz et al. where it was argued that fatal traffic accident risk increases by $\SI{6}{\percent}$ in the US due to Daylight Saving Time spring transition. This manuscript is a 1000 word correspondence showing that the bulk of this reported risk comes from the transition dates mandated by the Energy Policy Act in 2007.
\end{abstract}
\maketitle

\begin{widetext}
    This is the accepted manuscript of a correspondence published by Current Biology Volume 30, Issue 7, Pages R298-R300. ISSN 0960-9822, ESSN 1879-0445.

  For the authored, published version please visit doi \href{https://doi.org/10.1016/j.cub.2020.03.007}{https://doi.org/10.1016/j.cub.2020.03.007}

\end{widetext}

The impact of Daylight Saving Time (DST) transitions on the human circadian system and on everyday life is currently subject to close analysis.  \citet{Fritz2020} studied recently large scale United States (US) registry data (1996 to 2017) on fatal motor vehicle accidents (MVA) and reported the incidence rate ratio. The authors report results for data before 2006, for data after 2007, and for the whole observation period. They also report morning, afternoon and whole-day results. The discussion and conclusions are extracted mainly from the whole-day figures and those from the entire 1996 to 2017 period. Yet the breakdown illustrates a most interesting fact: the amendments in the Uniform Time Act made by the Energy Policy Act doubled the increase of fatal morning MVA attributed to the spring transition.

\begin{figure*}[tb]
  \centering
  
  \includegraphics[width=0.8\textwidth]{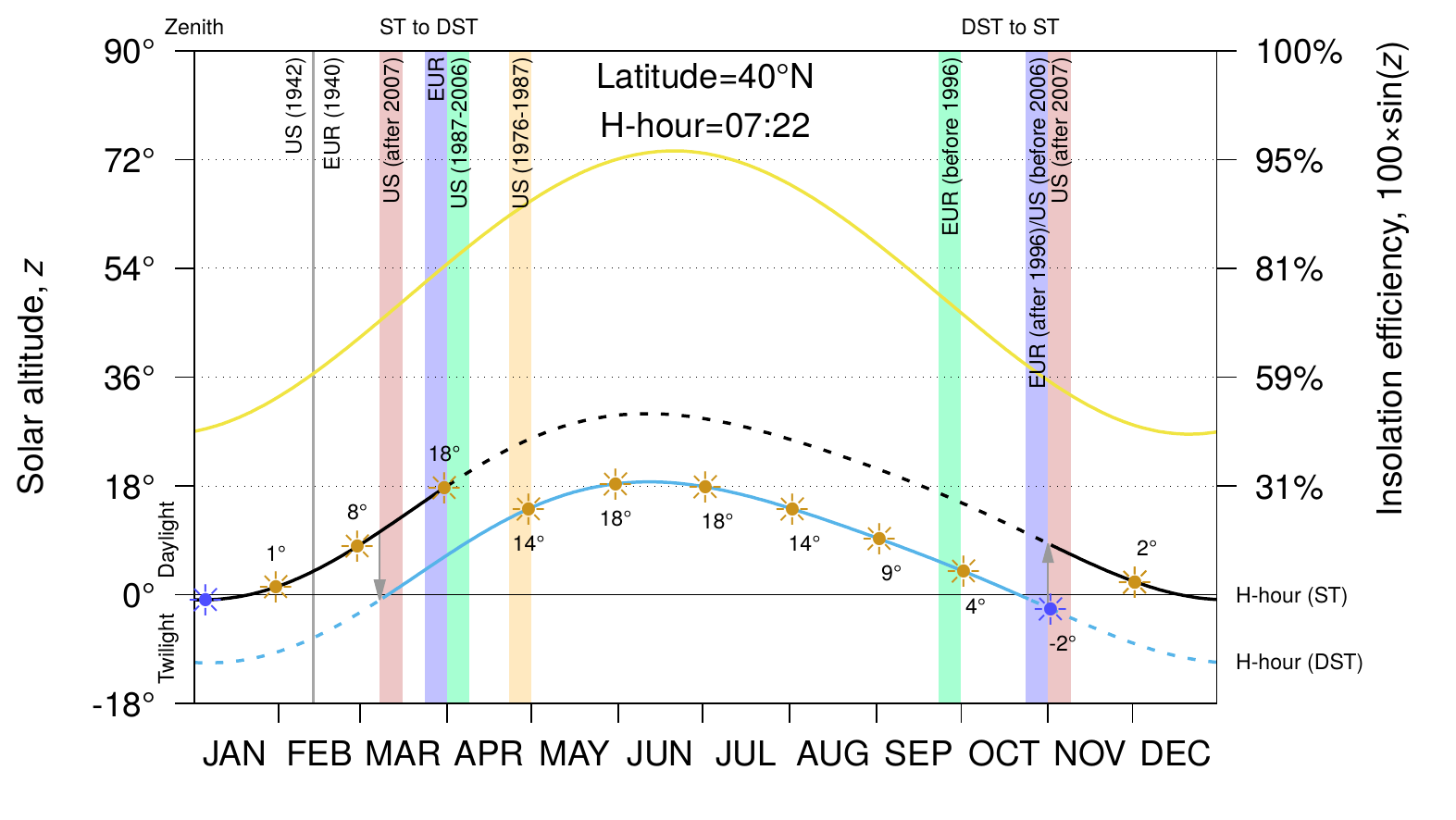}
  
  \caption{\textbf{The seasonal variation of solar altitude at \ang{40}N}. The solar altitude $z$ at the hour of the latest sunrise of the year ---black, H-hour (Standard Time, ST)--- and at the preceding hour --blueish, H-hour (Daylight Saving Time, DST)--- as a function of calendar date for \ang{40}N latitude where the H-hour is 07:22 mean solar time, the sunrise time on January the 5th. Coloured vertical strip bands display different arrangements of transition dates; each one expands for one week. The yellowish line displays solar altitude at solar noon. Sunrise is set at $z_c=\ang{-0.85}$ due to solar finite apparent size and atmospheric refraction.}
  \label{fig:angulo}
\end{figure*}

Figure~\ref{fig:angulo} shows a simplistic scenario at \ang{40}N ---the latitude of New York City and Madrid--- in which the yearly evolution of the solar altitude $z$ at the latest sunrise time ---here after the H-hour--- is plotted  (black line). Solar altitude is appropriate for understanding a key parameter related to the rate of traffic accidents: illumination conditions (see Figure 1B in Ref.~\cite{Fritz2020}). The H-hour is a function of latitude only if expressed as a mean solar time or as a distance to solar noon.  It impacts human social life since an activity starting at or after the H-hour would certainly occur in the photoperiod irrespective of calendar date. People at \ang{40}N latitude would see the Sun crossing the horizon in January 5th at the H-hour and would see the Sun as high as $z=\ang{30}$ above the horizon in June at the same hour of the day, if nothing else changes. As the Sun gets higher in the sky in spring/summer more people find natural to advance their activity aiming to mitigate their exposure to the highest insolation: at \ang{40}N insolation efficiency climbs up to a tropical \SI{95}{\percent} (see right axis in Figure~\ref{fig:angulo}). Likewise, as the solar altitude decreases in autumn/winter people delay their activity aiming to mitigate their exposure to morning darkness. This circanual cycle is observed in tropical, pre-industrial societies\cite{Yetish2015}. It is also observed in extratropical, industrial societies synchronized by clocks\cite{Monsivais2017a,Martin-Olalla2019f} and which have adopted year-round time schedules where DST transition dates regulate the mechanism. Chronobiologists largely criticizes\cite{Roenneberg2019,Fritz2020} the practice, yet they eventually acknowledge that people are prone to this natural behaviour\cite{Roenneberg2019a}.

In Figure~\ref{fig:angulo} the spring transition brings the human activity at the H-hour from the black line to the blue line. In fall, the shift is reversed. Vertical colored strip bands display transition dates. The Energy Policy Act advanced in 2007 the spring transition by three weeks. In 2020 $z$ will change from $z_{\text{st}}\sim\ang{12}$ to $z_{\text{dst}}\sim\ang{1}$ at \ang{40}N in the US. It changed from $z_{\text{st}}\sim\ang{19}$ to $z_{\text{dst}}\sim\ang{7}$ before 2006. In Europe at \ang{40}N, it will change from $z_{\text{st}}\sim\ang{17}$ to $z_{\text{dst}}\sim\ang{5}$ this year. 

Either of the last two cases keeps the Sun high enough over the horizon so that illumination conditions at the H-hour are unaltered. Therefore an increase of traffic accidents due to changing illumination conditions is expected to be slight. However, the changing in illumination conditions is dramatic for the current US regulations: a nice deal of traffic volume is shifted close to dawn after the transition.

Several cons and pros must be further considered: first, a fraction of population commutes before H-hour in the winter dawn, they are more prone to see the Sun below the horizon after DST begins in the middle of March; second, $z_d$ still worsens as latitude decreases (see Supplementary material); third, and although figure~\ref{fig:angulo} is showing solar properties and thus it is insensitive to longitude, human social activity is dictated by standard clock time which makes $z_{\text{dst}}$ worsen westward of a time meridian, an effect discussed by the authors. On the other side, human activity at the H-hour in the US slightly reduced compared to that in Europe at this time\cite{Martin-Olalla2018} which would allow a slightly larger extension of DST regulations in the US. 

\citet{Fritz2020}'s results remarkably agree with this framework. Figure 2 of the Ref.\cite{Fritz2020} attributes to DST spring transitions before 2006 and after 2007 a \SI{5}{\percent} increase in afternoon fatal MVAs and a \SI{6}{\percent} increase in morning fatal MVAs before 2006. The relevance of these results (~\SI{5}{\percent}) should be evaluated by comparing the outcome to the relative standard deviation (RSD) of weekly averaged fatal MVAs which is $\sim\SI{15}{\percent}$ (see Figure 1A of Ref.~\cite{Fritz2020}). The attributed increase is just one third of RSD. Also the lower endpoint of the confidence interval for each of these observations barely reaches 1.00–1.01. Understandably, the authors agree that this impact is slight. These observations can be plausibly linked to circadian misalignment and sleep deprivation since illumination conditions were unaltered here.

In sharp contrast, the morning increase attributed to DST after 2007 more than doubles the preceding results: \SI{13}{\percent} versus \SI{5}{\percent} to \SI{6}{\percent} with the lower endpoint of the confidence interval at 1.06, markedly greater than 1.00 (see Figure 2 of Ref.~\cite{Fritz2020}). Indeed the morning increase after 2007 is the one and only result close to RSD. It is therefore a marked impact. Following Figure~\ref{fig:angulo}, a plausible cause of this spike is the change in transition dates due to the Energy Policy Act, which brought a dramatic and predictable change in illumination conditions close to the rush hour. Notice that DST onset has regularly occurred in the Northern hemisphere from the end of March to the end of April, only after the sun is high enough in the sky.

Illumination issues could also arise at the fall transition. Currently $z_{\text{dst}}$ decreases to $\sim\ang{0}$ in Europe and to $\ang{-2}$ in the US ---meaning that the latest sunrise (clock) time of the year actually happens on the first Saturday of November--- compared to $z_{\text{dst}}\sim\ang{5}$ on the last Sunday of September, a transition date in continental Europe discontinued in 1996. Very likely an advance of the fall transition date by some three weeks could result in a decrease of morning MVAs, mimicking the results obtained by Fritz et al.

Summarizing, the analysis shows that the acute impact of DST transitions on incidence rate ratio for weekly average fatal MVAs is close to \SI{5}{\percent}, one third of the RSD of observations, a slight impact. However, the results show that the Energy Policy Act more than doubled morning incidence rate ratio up to \SI{13}{\percent} due to a marked change in the illumination conditions. Based on the first observation, the authors call for discontinuing DST policies, despite its impact being slight. Instead the outstanding analysis provided by Fritz et al. calls for discontinuing the provisions of the Energy Policy Act relative to transition dates.

\acknowledgments

Prof. Dr. Juan C. del Álamo from University of Washington and Prof. Dr. Manuel García-Villalba from Universidad Carlos III alerted JMM-O of the commented paper. JMM-O thanks Prof. Dr. Jorge Mira from Universidade de Santiago de Compostela for a critical reading of this letter. An earlier version of Figure~\ref{fig:angulo} was presented in a meeting held on 2018 at the Consello da Cultura Galega in Santiago de Compostela (Spain). The figure was produced by \texttt{gnuplot}. It plots a file produced by a script in \texttt{octave} which uses \texttt{xplanet} to compute the solar declination in the year 2020 and a script by Darin C. Koblick ---available at \href{https://www.mathworks.com/matlabcentral/fileexchange/32793-equation-of-time}{https://www.mathworks.com/matlabcentral/fileexchange/32793-equation-of-time}--- to compute the equation of time in 2020 from which sunrise times and solar altitudes were produced.

This project was started on February 1st, 2020. It was not funded.

\end{document}